\documentclass[twocolumn,showpacs,preprintnumbers,amsmath,amssymb]{revtex4}
\date{March 2010}
\usepackage{epsfig}
\usepackage{latexsym}
\usepackage{amssymb}
\usepackage{booktabs}
\usepackage{graphicx}% Include figure files
\newcommand{\be}{\begin{equation}}
\newcommand{\ee}{\end{equation}}
\newcommand{\ba}{\begin{eqnarray}}
\newcommand{\ea}{\end{eqnarray}}
\newcommand{\bi}{\begin{itemize}}
\newcommand{\ei}{\end{itemize}}

\newcommand{\half}{{\textstyle\frac{1}{2}}}

\renewcommand{\>}{\rangle}

\newcommand{\la}{\label}

\newcommand{\amuhvp}{a_\mu^{\rm hvp}}
\newcommand{\mmu}{m_\mu}

\newcommand{\fs}{f^{\rm sub}}
\newcommand{\Qr}{Q_{\rm ref}}

\begin{document}
% \preprint{MITP/17-XXX~~~~ HIM-2017-06}
\title{Lorentz-covariant coordinate-space representation of the 
leading hadronic contribution to the anomalous magnetic moment of the muon}
% \title{Lattice QCD calculations of the hadronic vacuum polarization: methodological aspects}

\author{Harvey~B.~Meyer}

\affiliation{\vspace{0.2cm}PRISMA Cluster of Excellence,
Institut f\"ur Kernphysik and Helmholtz~Institut~Mainz,
Johannes Gutenberg-Universit\"at Mainz,
D-55099 Mainz, Germany\vspace{0.2cm}}

\date{\today}

\begin{abstract}
We present a Lorentz-covariant, Euclidean coordinate-space
expression for the hadronic vacuum polarisation, the Adler function and the 
leading hadronic contribution to the anomalous magnetic moment of the muon. 
The representation offers a lot of flexibility for an implementation in lattice QCD.
We expect it to be particularly helpful for the quark-line disconnected contributions.
\end{abstract}

% \pacs{12.38.Gc, 12.38.Mh, 25.75.-q}
 % 13.40.Gp
\maketitle

\section{Introduction}

Two-point functions of the quark-flavor currents $\bar q\gamma_\mu q$
play an importance role in precision tests of the Standard Model of
particle physics. The electromagnetic current $j_\mu^{\rm em}$
correlator enters the running of the QED coupling constant, and the
non-diagonal correlation of $j_\mu^{\rm em}$ with the weak-isospin
current $j_\mu^3$ contributes to the running of the weak mixing
angle~\cite{Jegerlehner:1985gq}.  Furthermore, the leading hadronic
contribution to the anomalous magnetic moment of the muon $\amuhvp$
can be determined by non-perturbative theory methods from the
electromagnetic current correlator. The contribution $\amuhvp$, where
`HVP' stands for hadronic vacuum polarization, represents the largest
uncertainty in the Standard Model prediction for this precision
observable.  Given the experiments in preparation~\cite{Eads:2015arb}
at FermiLab~\cite{E-989web} and JPARC~\cite{JPARCg-2web}, which are
expected to improve the accuracy of the direct measurement by a factor
4, it is important to reduce the uncertainty on the prediction by a
comparable factor.  While the phenomenological determination of
$\amuhvp$ via its dispersive representation is still the most accurate
approach~\cite{Jegerlehner:2009ry,Blum:2013xva}, a purely theoretical prediction is both conceptually
desirable and provides for an independent check. Since the
vacuum polarization is inserted into an integral which is strongly
weighted to the low-energy domain, calculating the hadronic vacuum
polarization has become an important goal for several lattice QCD
collaborations 
\cite{Blum:2002ii,Gockeler:2003cw,Aubin:2006xv,Feng:2011zk,Boyle:2011hu,DellaMorte:2011aa,Burger:2013jya,
Blum:2015you,Blum:2016xpd,Chakraborty:2016mwy,Borsanyi:2016lpl,DellaMorte:2017dyu}.

In lattice QCD, which representation of the $\amuhvp$ is used matters,
since the choice affects the systematic and statistical uncertainties
of the result.  In the pioneering article~\cite{Blum:2002ii},
$\amuhvp$ was written as an integral of the vacuum polarization
$\Pi(Q^2)$ over all virtualities $Q^2$.  Later, other representations
were proposed, in particular the time-moment
representation~\cite{Bernecker:2011gh,Feng:2013xsa,Francis:2013fzp},
where the vector correlation function is projected onto vanishing
spatial momentum and integrated as a function of Euclidean time.  In
this article we present a manifestly Lorentz-invariant representation
of the vacuum polarization $\Pi(Q^2)$, the Adler function $Q^2\Pi'(Q^2)$ and
$\amuhvp$ based on the coordinate-space representation of the vector
correlator. We call it the covariant coordinate-space (CCS)
representation.  Besides its formal elegance, we expect it to be
helpful in lattice QCD calculations, as the Lorentz symmetry 
present in the continuum
leads to a lot of flexibility and opportunities for cross-checks in its
implementation.  We think it will be especially beneficial for
disconnected-diagram contributions~\cite{Francis:2014hoa,Guelpers:2015nfb,Blum:2015you,Kawanai:2017gdd}, 
where, in the standard algorithm,
the noise-to-signal ratio on the coordinate-space correlator increases
rapidly at long distances.

The structure of this paper is as follows.  Our basic definitions and
the derivation of the position-space expressions are presented 
in the next section. A test and illustration of the method is also provided. 
Section \ref{sec:LQCD} is devoted to some aspects
of the implementation of the method in lattice QCD.

% Some numerical tests in lattice perturbation theory are described in
% section \ref{sec:numerics} and the results are given in section \ref{sec:results}.

\renewcommand{\vec}[1]{\boldsymbol{#1}}

\section{Derivation of the covariant coordinate-space expressions}

In this section we consider continuum QCD in infinite Euclidean space.  The
conserved vector current is defined as
$j_\mu(x)=\bar\psi(x)\gamma_\mu\psi(x)$, where the Dirac matrices are
all hermitian and satisfy
$\{\gamma_\mu,\gamma_\nu\}=2\delta_{\mu\nu}$.  

\subsection{Definitions}

 The primary object is the position-space correlator
\be
G_{\mu\nu}(x) = \<j_\mu(x) j_\nu(0)\>.
\ee
The polarization tensor is its Fourier transform,
\be\la{eq:PolTens}
\Pi_{\mu\nu}(Q) \equiv \int d^4x \, e^{iQ\cdot x} G_{\mu\nu}(x),
\ee
and O(4) invariance and current conservation imply the tensor structure
\be\la{eq:PimunuQ}
\Pi_{\mu\nu}(Q) = \big(Q_\mu Q_\nu -\delta_{\mu\nu}Q^2\big) \Pi(Q^2).
\ee
With these conventions, the spectral function
\be\la{eq:rhoGG}
\rho(q^2) \equiv -\frac{1}{\pi} {\rm Im} \Pi(Q^2)\Big|_{{Q_0=-iq_0+\epsilon},\,{\vec Q=\vec q}} 
\ee
is non-negative for a flavor-diagonal correlator. 
For the electromagnetic current, it is related to the $R$ ratio via
\be \la{eq:rhoR}
\rho(s) =\frac{R(s)}{12\pi^2},
\qquad
R(s) \equiv  \frac{\sigma(e^+e^-\to {\rm hadrons})}
 {4\pi \alpha(s)^2 / (3s) } .
\ee
The denominator is the treelevel cross-section $\sigma(e^+e^-\to\mu^+\mu^-)$
in the limit $s\gg m_\mu^2$, and we have neglected QED corrections.
The vacuum polarization, and the Adler function
\be
{\cal A}(Q^2)\equiv Q^2\frac{d}{dQ^2} \Pi(Q^2),
\ee
are recovered through a dispersion relation,
\ba \la{eq:DispRel}
 \Pi(Q^2)-\Pi(0) &=& Q^2 \int_0^\infty ds \frac{\rho(s)}{s(s+Q^2)},
\\
{\cal A}(Q^2) &=& Q^2 \int_0^\infty ds \frac{\rho(s)}{(s+Q^2)^2}.
\la{eq:DispRelA}
\ea

\subsection{Derivation}
As a motivation, we start from the expression~\cite{Knecht:2003kc} for $\amuhvp$ in terms of the Adler function,
\be
\amuhvp = 2\pi^2 \left(\frac{\alpha}{\pi}\right)^2 \int_0^1 \frac{dy}{y}(1-y)(2-y)\, {\cal A}(Q^2(y)),
\ee
where 
\be
Q^2(y) = \frac{y^2}{1-y}\, m_\mu^2 \quad \leftrightarrow \quad 
y = \frac{2|Q|}{|Q|+\sqrt{4m_\mu^2+Q^2}}.
\ee
% Performing the change of variables back 
Returning to the integration variable $Q^2$, we obtain
\ba\la{eq:amuAdler}
\amuhvp &=& \int_0^\infty dQ^2\; {\cal A}(Q^2) \; g_a(Q^2),
\\
g_a(Q^2) &=& 2\alpha^2 \frac{\mmu^4}{|Q|^6}\,y(|Q|)^4.
\ea
However, we will keep the derivation more general and consider a general Lorentz-scalar physical quantity
derived from the vector correlator,
\be\la{eq:Phi}
\Phi[g]  = \int_0^\infty dQ^2\; {\cal A}(Q^2) \; g(Q^2),
\ee
for some function $g(Q^2)$. 
Below, we reinterpret Eq.\ (\ref{eq:Phi}) as a four-dimensional integral with 
spherical symmetry\footnote{The unit sphere in four dimensions has a surface of $2\pi^2$.},
\be\la{eq:PhiSpher}
\Phi[g]= \frac{1}{\pi^2}\int \frac{d^4Q}{Q^2}\; {\cal A}(Q^2)\; g(Q^2).
\ee

We project out the transverse component of the polarisation tensor,
\be
\pi_T(Q) = \Big(\delta_{\mu\nu} - \frac{Q_\mu Q_\nu}{Q^2}\Big) \Pi_{\mu\nu}(Q)
= -3Q^2 \Pi(Q^2).
\ee
The Adler function can be expressed via $\pi_T(Q)$ via
\ba
{\cal A}(Q^2) &=& \frac{1}{3Q^2} \Big[\pi_T(Q) - \frac{Q_\lambda}{2} \frac{\partial}{\partial Q_\lambda} \pi_T(Q)\Big]
\ea
Inserting the position-space correlator,
\be
\pi_T(Q) = \Big(\delta_{\mu\nu} - \frac{Q_\mu Q_\nu}{Q^2}\Big) \int d^4x\; G_{\mu\nu}(x) \;e^{iQ\cdot x},
\ee
we obtain 
\ba
{\cal A}(Q^2) &=& \frac{1}{3Q^2}\Big(\delta_{\mu\nu} - \frac{Q_\mu Q_\nu}{Q^2}\Big)\times
\\ && \times\int d^4x\;G_{\mu\nu}(x)\; e^{iQ\cdot x}\; (1- \frac{i}{2}(Q\cdot x)).
\nonumber
\ea
Inserting this expression into Eq.\ (\ref{eq:PhiSpher}) and interchanging the order of the momentum-space
and position-space integrals, one reaches
\ba\la{eq:master1}
\Phi[g] &=& \int d^4x\;G_{\mu\nu}(x)\; H_{\mu\nu}(x),
% \frac{1}{3\pi^2}(1- \frac{x_\lambda}{2}\frac{\partial}{\partial x_\lambda}) I_{\mu\nu}(x),\qquad 
\\
H_{\mu\nu}(x) &=& \frac{1}{3\pi^2}(1- \frac{x_\lambda}{2}\frac{\partial}{\partial x_\lambda}) I_{\mu\nu}(x),\qquad 
\\
I_{\mu\nu}(x) &=& \int \frac{d^4Q}{(Q^2)^2} \,g(Q^2) \Big(\delta_{\mu\nu} - \frac{Q_\mu Q_\nu}{Q^2}\Big)\, e^{iQ\cdot x}.
\ea
The kernel $I_{\mu\nu}(x)$ can be expressed as 
\ba\la{eq:ImunuI}
I_{\mu\nu}(x) &=& (\partial_\mu^{(x)}\partial_\nu^{(x)} - \delta_{\mu\nu} \triangle_x) I(x^2),
\\
I(x^2) &=& \int\frac{d^4Q}{(Q^2)^3}\; g(Q^2)\; e^{iQ\cdot x}
\\ &=& \frac{4\pi^2}{|x|}\int_0^\infty \frac{d|Q|}{|Q|^4}\,g(Q^2)\, J_1(|Q||x|),
\ea
where we performed the angular integration of the Fourier transform and the $J_n(z)$ are the Bessel functions of the first kind.
Using the chain rule, we obtain
\ba\la{eq:ImunuII}
I_{\mu\nu}(x) &=& -\delta_{\mu\nu}\Big(\frac{\partial^2I}{\partial|x|^2} + \frac{2}{|x|} \frac{\partial I}{\partial |x|}\Big)
\\ && + \frac{x_\mu x_\nu }{x^2} \Big(\frac{\partial^2I}{\partial|x|^2} - \frac{1}{|x|} \frac{\partial I}{\partial |x|} \Big).
\nonumber
\ea
and 
\ba
&& (1-\frac{x_\lambda}{2}\frac{\partial}{\partial x_\lambda})I_{\mu\nu}(x)
 = - \delta_{\mu\nu}\Big(-\frac{|x|}{2} \frac{\partial^3I}{\partial |x|^3} + \frac{3}{|x|} \frac{\partial I}{\partial |x|}\Big)
\nonumber\\ && + \frac{x_\mu x_\nu}{x^2}\Big(-\frac{|x|}{2} \frac{\partial^3I}{\partial |x|^3} + \frac{3}{2} \frac{\partial^2I}{\partial |x|^2} 
- \frac{3}{2|x|} \frac{\partial I}{\partial |x|}\Big).
\ea
We note that an $x$-independent term in $H_{\mu\nu}(x)$ would not contribute to $\amuhvp$, since
\be
\int d^4x \;G_{\mu\nu}(x) = 0\quad \forall \mu,\nu
\ee
in the vacuum, when all correlation lengths are finite. We make use of this property to subtract from $H_{\mu\nu}(x)$ 
an $x$-independent term proportional to $\delta_{\mu\nu}$.
The derivatives with respect to $|x|$ act on $J_1(|Q||x|)/|x|$, resulting in Bessel functions of higher order. We define
\ba
h_1(z) &=& 
\frac{3}{8} + \Big(-\frac{z}{2} \frac{\partial^3}{\partial z^3} + \frac{3}{z} \frac{\partial }{\partial z}\Big) \frac{J_1(z)}{z}
\\ &=& \frac{3}{8} + \frac{1}{2}J_0(z) - \frac{5}{2} \frac{J_1(z)}{z} + 3 \frac{J_2(z)}{z^2},
\\
h_2(z) &=& \Big(-\frac{z}{2} \frac{\partial^3}{\partial z^3} + \frac{3}{2} \frac{\partial^2}{\partial z^2} 
- \frac{3}{2z} \frac{\partial }{\partial z}\Big) \frac{J_1(z)}{z}
\\ &=& \frac{1}{2z^3}\Big(z(z^2-24)J_0(z) - 8(z^2-6) J_1(z) \Big).\quad 
\ea
We note that for $z\to0$, $h_i(z)\sim \lambda_i z^4$ with 
\be\la{eq:lambdai}
\lambda_1= \frac{7}{3072},\qquad \lambda_2= \frac{1}{768}.
\ee
Thus, inserting the explicit form of $g(Q^2)$, we obtain finally
the physical quantity of interest $\Phi[g] = \int d^4x\;G_{\mu\nu}(x)\; H_{\mu\nu}(x)$ with 
\ba
\la{eq:master2}
% \Phi[g] &=& \int d^4x\;\<j_\mu(x)j_\nu(0)\> \times
% \\ && \times\Big(-\delta_{\mu\nu} {\cal H}_1(|x|) + \frac{x_\mu x_\nu}{x^2} \,{\cal H}_2(|x|)\Big),
H_{\mu\nu}(x) &=& -\delta_{\mu\nu} {\cal H}_1(|x|) + \frac{x_\mu x_\nu}{x^2} \,{\cal H}_2(|x|),
\nonumber\\
{\cal H}_i(|x|) &=& \frac{2}{3}\int_0^\infty \frac{dQ^2}{Q^2}\; h_i(|Q||x|)\, g(Q^2).
\ea
Thus, for Eq.\ (\ref{eq:master1}) to provide an explicit expression for $\Phi[g]$, it suffices to pre-compute the weight functions ${\cal H}_i(|x|)$.
For now we note that once the spacetime indices of $G_{\mu\nu}(x)$ and $H_{\mu\nu}(x)$ 
are contracted, the integral $\int d^4x\to 2\pi^2 \int_0^\infty d|x|\,|x|^3$ 
becomes one-dimensional. Secondly, since  $I_{\mu\nu}(x)$ and $H_{\mu\nu}(x)$ are transverse tensors, 
$\partial_\mu^{(x)}H_{\mu\nu}(x)=0$, a relation exists between the weight functions,
\be\la{eq:F1F2rel}
{\cal H}_1^{\,\prime}(|x|)= {\cal H}_2^{\,\prime}(|x|) + \frac{3{\cal H}_2(|x|)}{|x|}.
\ee

\subsubsection{Computing the Adler function}

To obtain the Adler function, $\Phi[g]\doteq {\cal A}(Q^2)$, we simply set 
\be
g(\Qr^2)\doteq \delta(Q^2-\Qr^2),
\ee
and obtain immediately
\be
{\cal H}_i(|x|) = \frac{2\,h_i(|Q||x|)}{3Q^2}\,.
\ee
A particularly elegant expression results in the limit $Q^2\to 0$
for the slope of the Adler function, or equivalently of the vacuum polarisation,
\ba\la{eq:Ap0}
&& {\cal A}'(0)= \Pi'(0)
\\ && = \frac{1}{1152} \int d^4x\;G_{\mu\nu}(x)\, (x^2)^2 
\Big(-\frac{7}{4}\delta_{\mu\nu} + \frac{x_\mu x_\nu}{x^2}\Big).
\nonumber
\ea
Since it is well-known that the leading hadronic contribution to the anomalous magnetic moment of 
a lepton in the massless-lepton  limit is given by the slope of the Adler function at the origin,
\be\la{eq:aeAp0}
\lim_{\mmu\rightarrow 0}\frac{\amuhvp}{\mmu^2} = \frac{4}{3}\alpha^2  {\cal A}'(0),
\ee
Eq.\ (\ref{eq:Ap0}) can be used to calculate the hadronic contribution to the $(g-2)$ of the electron.

\subsubsection{The subtracted vacuum polarisation}

To obtain the subtracted vacuum polarisation,
\be
\Phi[g]\doteq \Pi(Q^2)-\Pi(0),
\ee
we set
\be
g(\Qr^2)\doteq \frac{1}{\Qr^2}\,\theta(Q^2-\Qr^2),
\ee
and obtain ${\cal H}_i(|x|) = x^2\cdot \bar{\cal H}_i(|Q||x|)$, where 
\ba
 \bar{\cal H}_1(z) &=&  \frac{z^2}{4608} \Big\{ 24 \,
   _2F_3\left(1,1;2,3,3;-\frac{z^2}{4}\right)
\\ && -20 \,  _2F_3\left(1,1;2,3,4;-\frac{z^2}{4}\right)
\nonumber\\ && +3 \,   _2F_3\left(1,1;2,3,5;-\frac{z^2}{4}\right)\Big\}
\nonumber
\ea
and
\ba
 \bar{\cal H}_2(z) &=& \frac{z^2}{1152} \Big\{6 \,
   _2F_3\left(1,1;2,3,3;-\frac{z^2}{4}\right)
\\ &&  -8 \,   _2F_3\left(1,1;2,3,4;-\frac{z^2}{4}\right)
\nonumber\\ && +4 \,   _2F_3\left(1,1;2,4,4;-\frac{z^2}{4}\right)
\nonumber\\ && -\,   _2F_3\left(1,1;2,4,5;-\frac{z^2}{4}\right)\Big\}.
\nonumber
\ea
The functions in the brackets are the generalized hypergeometric functions.
% \footnote{In Mathematica notation, for instance $_2F_3\left(1,1;2,3,3;-\frac{z^2}{4}\right)=
% {\rm HypergeometricPFQ}[\{1, 1\}, \{2, 3, 3\}, -(z^2/4)]$.}

\subsubsection{The case of $\amuhvp$}

Here we set $g(Q^2)=g_a(Q^2)$, and the weight functions can be written
\ba\la{eq:main}
{\cal H}_i(|x|) &=& \frac{8\alpha^2}{3\mmu^2}f_i(\mmu|x|),
\\
\la{eq:fi1}
f_i(\mmu|x|) &=& \mmu^6 \int_0^\infty \frac{d|Q|}{|Q|^3}\, \left(\frac{y(|Q|)}{|Q|}\right)^4\; h_i(|Q||x|).
\qquad
\ea

To study the dimensionless weight functions $f_i(z)$, we perform the change of variables $\bar Q = |Q|/\mmu$,
\be\la{eq:fi2}
f_i(z) = \int_0^\infty \frac{d\bar Q}{\bar Q^3}\; \frac{16 h_i(z\,\bar Q)}{(\bar Q+\sqrt{4+\bar Q^2})^4}\; .
\quad
\ee
Note that since the weight functions  only depend on $\mmu |x|$,  the small $|x|$ behavior is at the same time
the small $\mmu$ behavior.
Making use of the integral $\int_0^\infty dv\, \frac{16v}{(v+\sqrt{4+v^2})^4}=\frac{1}{3}$, we obtain
\be\la{eq:fi_orig}
f_i(z) \stackrel{z\to0}{=} \frac{\lambda_i}{3} \, z^4.
\ee
Thus in the limit of vanishing lepton mass, we recover Eqs.\ (\ref{eq:aeAp0}--\ref{eq:Ap0}).

The regime of large $|x|$ is more easily investigated from Eq.\ (\ref{eq:fi1}) by performing the change of variables
$v = |Q||x|$,
\be\la{eq:fi3}
f_i(z) = z^6  \int_0^\infty \frac{dv}{v^3} \frac{16\,h_i(v)}{(v+\sqrt{4z^2+v^2})^4}\, .
\ee
Taking the limit of large argument (corresponding to large $|x|$), $f_i(z) \stackrel{z\to\infty}{=} \nu_i \,z^2$,
with
\ba
\nu_i &=& \int_0^\infty \frac{dv}{v^3}  h_i(v),
\quad
\nu_1 = \frac{5}{192}, \quad \nu_2 = \frac{1}{96}.
\ea
By expanding the denominator of $f_2(z)$ further, one obtains the first terms of a series
in $1/z$,
\be\la{eq:f2largez}
f_2(z) = \frac{z^2}{96} - \frac{z}{15}+ \frac{1}{4} - \frac{5}{8z} + \dots
\ee
The subleading terms cannot directly be obtained from the representation (\ref{eq:fi3}).

We now derive an expression for the weight function in terms of known special functions.
From Eq.\ (\ref{eq:F1F2rel}), since both weight functions are proportional to $z^4$ at small $z$, the integration yields
\be\la{eq:f1Integ}
f_1(z) = f_2(z) + 3\int_0^z d\bar z\,\frac{f_2(\bar z)}{\bar z}.
\ee
We will therefore first address $f_2(z)$, and then obtain $f_1(z)$ from this equation.

\begin{figure}[t]
\includegraphics[width=.5\textwidth]{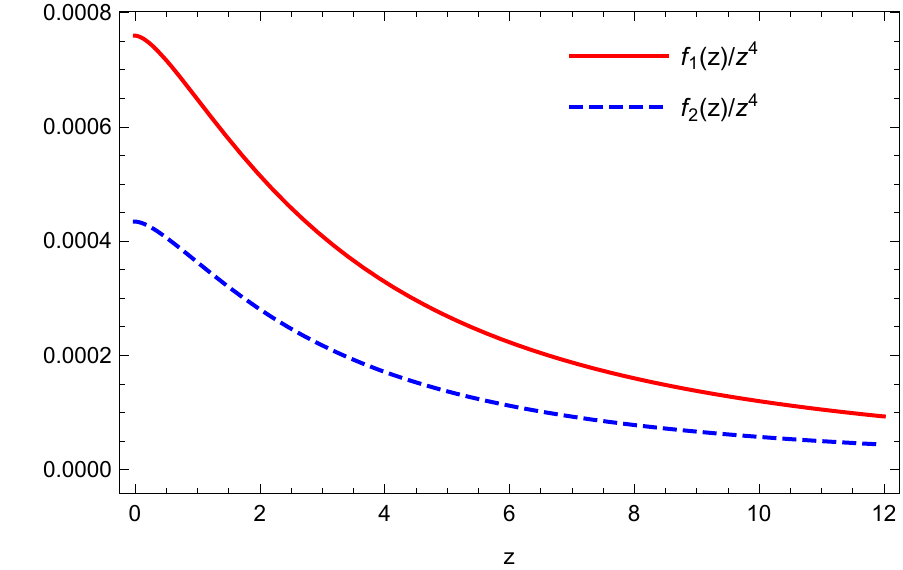}
\caption{The weight functions $f_1(z)$ and $f_2(z)$ needed to calculate $\amuhvp$ using 
Eqs.\ (\ref{eq:master1}, \ref{eq:master2}, \ref{eq:main}, \ref{eq:f1final}, \ref{eq:f2final}).
In the calculation of $\amuhvp$, the argument of the weight functions is $(\mmu|x|)$,
so that $z=1$ corresponds to 1.87\,fm. Note that $\lim_{z\to0} f_1(z)/z^4=\frac{7}{9216}$ and $\lim_{z\to0} f_1(z)/z^4=\frac{1}{2304}$;
see Eqs.\ (\ref{eq:lambdai}) and (\ref{eq:fi_orig}).}
\label{fig:fi}
\end{figure}

Suppose that we want to calculate a function $\phi(z)$, and that we are initially able 
to calculate the expression
\be
-L_r \phi(r) \equiv 
-\Big(\frac{d^2}{dr^2} + \frac{3}{r} \frac{d}{dr}\Big) \phi(r) = \rho(r).
\ee
The differential operator $L_r$ appearing in this equation is nothing but the four-dimensional Laplacian
in spherical coordinates, $r$ playing the role of the radial coordinate. We are then dealing 
with an electrostatic problem in 4+1 dimensions for a spherically symmetric charge distribution.
Therefore, assuming $\rho(r)$ falls off faster than $1/r^2$ at large $r$, the function
$\phi(r)$ can be obtained using the Green's function $G_0(x) = \frac{1}{4\pi^2 x^2}$,
\ba
\phi(|x|) &=& \int d^4y \;G_0(x-y)\; \rho(|y|).
\ea
Defining
\ba
d_0(|x|,|y|) &=& \theta(x^2-y^2) \frac{1}{|x|^{2}} + \theta(y^2-x^2) \frac{1}{|y|^{2}},
\qquad
\ea
the convolution integral can be simplified using Gauss's theorem and exploiting the spherical symmetry
of $\rho(r)$,
\ba\la{eq:phiconvol}
\phi(|x|) &=& \frac{1}{2} \int_0^\infty d|y|\,|y|^3 \,d_0(|x|,|y|)\,\rho(|y|)
\\ &=& \frac{1}{2|x|^2}\int_0^{|x|}d|y|\;|y|^3 \,\rho(|y|)
+ \frac{1}{2} \int_{|x|}^\infty dy\,|y|\,\rho(|y|).
\nonumber
\ea
We start from Eq.\ (\ref{eq:fi2}), note that 
\be
% \Big(\frac{d^2}{dz^2} + \frac{3}{z} \frac{d}{dz}\Big) 
L_z (z^2 h_2(z))= \frac{z^2}{2}J_2(z),
\ee
and compute the integral
\ba\la{eq:Lz2f2}
&&  % \Big(\frac{d^2}{dz^2} + \frac{3}{z} \frac{d}{dz}\Big)
L_z (z^2f_2(z)) = \frac{z^2}{2}\int_0^\infty \frac{d\bar Q}{\bar Q}\; \frac{16\,J_2(\bar Q z)}{(\bar Q+\sqrt{4+\bar Q^2})^4} 
\qquad 
% \\ && = \frac{z^2}{2\sqrt{\pi}}{\M}[\{\{\frac{1}{2},1\},\{\}\},\{\{1,2\},\{-2,-1\}\},z^2].\nonumber
\\ && = \frac{z^2}{2 \sqrt{\pi }}  G_{2,4}^{2,2}\left(z^2|
\begin{array}{c}
 \frac{1}{2},1 \\
 1,2,-2,-1 \\
\end{array}
\right),
\ea
where $G^{m,n}_{p,q}$ represents Meijer's function.
In order to recover $f_2(z)$ itself, we would like to apply Eq.\ (\ref{eq:phiconvol}).
However, $z^2f_2(z)\sim z^4$ at large $z$. We therefore perform subtractions using (\ref{eq:f2largez}),
\be\la{eq:f2sub}
\fs_2(z) \equiv f_2(z) - \Big( \frac{z^2}{96} - \frac{z}{15}+ \frac{1}{4} - \frac{5}{8z} \Big),
\ee
which obeys
\be\la{eq:mLz2fs}
-L_z (z^2\fs_2(z)) = -L_z (z^2f_2(z)) + \frac{z^2}{4}-z+2-\frac{15}{8z}.
\ee
We thus obtain the representation
\be\la{eq:fsInt}
z^2 \fs_2(z) = \frac{1}{2}\int_0^\infty d\bar z \,\bar z^3 d_0(z,\bar z) (-L_{\bar z}(\bar z^2 \fs_2(\bar z))).
\ee
After expression (\ref{eq:Lz2f2}) is inserted into Eq.\ (\ref{eq:mLz2fs}) and the latter in turn into Eq.\ (\ref{eq:fsInt}),
$\fs_2(z)$ can be expressed again in terms of Meijer's function. Reintroducing the subtracted terms from (\ref{eq:f2sub}),
we finally obtain 
\ba\la{eq:f2final}
&& f_2(z) =
\frac{G_{2,4}^{2,2}\left(z^2|
\begin{array}{c}
 \frac{7}{2},4 \\
 4,5,1,1 \\
\end{array}
\right)-G_{2,4}^{2,2}\left(z^2|
\begin{array}{c}
 \frac{7}{2},4 \\
 4,5,0,2 \\
\end{array}
\right)}{8 \sqrt{\pi } z^4}.
\qquad
% \\ && = \frac{1}{8\sqrt{\pi}z^4}\cdot
% \\ && \cdot \Big( \M[\{\{\frac{7}{2},4\},\{\}\},\{\{4,5\},\{1,1\}\},z^2]
% \nonumber\\ && - \M[\{\{\frac{7}{2},4\},\{\}\},\{\{4,5\},\{0,2\}\},z^2]\Big).
% \nonumber
\ea
From Eq.\ (\ref{eq:f1Integ}), we then obtain
\ba\la{eq:f1final}
 f_1(z) &=& f_2(z)
-\frac{3}{16 \sqrt{\pi }}\cdot
% \\ && \cdot
 \bigg[G_{3,5}^{2,3}\left(z^2|
\begin{array}{c}
 1,\frac{3}{2},2 \\
 2,3,-2,0,0 \\
\end{array}
\right) \qquad 
\\ && -G_{3,5}^{2,3}\left(z^2|
\begin{array}{c}
 1,\frac{3}{2},2 \\
 2,3,-1,-1,0 \\
\end{array}
\right)\bigg]\qquad 
\nonumber
% \\ && = f_2(z)
%  - \frac{3}{16\sqrt{\pi}} \cdot 
% \\ &&  \cdot \Big(
%    \M[\{\{1,\frac{3}{2},2\},\{\}\},\{\{2,3\},\{-2,0,0\}\},z^2] 
% \nonumber
% \\ &&  - \M[\{\{1,\frac{3}{2},2\},\{\}\},\{\{2,3\},\{-1,-1,0\}\},z^2]\Big).
% \nonumber
\ea
The functions $f_i(z)/z^4$ are displayed in Fig.\ \ref{fig:fi}.
The latter ratios depend very weakly on z in the range where they will be needed
to compute $\amuhvp$.

\subsection{Test of the CCS method}

To test the method, we construct a simple model for the position-space correlator $G_{\mu\nu}(x)$
based on its spectral representation. The latter reads
\ba\la{eq:jmujnuSpecRep}
G_{\mu\nu}(x) &=& (\partial_\mu^{(x)}\partial_\nu^{(x)} - \delta_{\mu\nu} \triangle_x) 
\tilde \Pi(|x|),
\\
\tilde \Pi(|x|) &=& \int_0^\infty ds\,\rho(s)\, G_{\sqrt{s}}(x),
\ea
where 
\be
G_m(x) = \frac{m}{4\pi^2 |x|}K_1(m|x|)
\ee
is the scalar propagator. The derivatives in Eq.\ (\ref{eq:jmujnuSpecRep})
act like in Eq.\ (\ref{eq:ImunuI}), see Eq.\ (\ref{eq:ImunuII}).
For illustration and testing purposes, we choose as a model
\be\la{eq:rhomodel}
\rho(s) = \frac{2}{3}f_V^2 M^2 \delta(s-M^2).
\ee
We note that the physical quantities to be computed are linear functions of the spectral
function $\rho(s)$, and in a finite-volume system the most general spectral function is a linear
combination of contributions of the type (\ref{eq:rhomodel}).
We remark that in the latter equation, the decay constant $f_V$ is dimensionless.
Then, with $r=|x|$, we obtain 
\ba\la{eq:GmunuModel}
&& G_{\mu\nu}(x) = 
\\ && \frac{f_V^2 M^3}{6\pi^2} \Big[-\delta_{\mu\nu} \frac{M}{r^2}  \Big(K_2(Mr) + Mr K_1(Mr) \Big)
\nonumber\\ &&  + \frac{x_\mu x_\nu}{x^2} \frac{1}{r^3}\Big(4Mr K_0(Mr) + (M^2r^2+8)K_1(Mr) \Big)\Big].
\nonumber
\ea
In general, if the position-space correlator is written in the form
\be
G_{\mu\nu}(x) = -\delta_{\mu\nu}\,{\cal G}_1(|x|) + \frac{x_\mu x_\nu}{x^2} {\cal G}_2(|x|),
\ee
physical quantities derived from it can be obtained from the scalar integral
\be\la{eq:GHred}
\Phi[g] = 2\pi^2 \int_0^\infty dr\,r^3\,\Big[{\cal H}_1(4{\cal G}_1-{\cal G}_2) + {\cal H}_2({\cal G}_2-{\cal G}_1)\Big].
\ee
Using Eq.\ (\ref{eq:Ap0}), one then finds
\be
{\cal A}'(0) = \frac{2}{3} \frac{f_V^2}{M^2}.
\ee
This matches the value obtained directly from Eq.\ (\ref{eq:DispRelA}),
\be
{\cal A}'(0) = \int_0^\infty ds\;\frac{\rho(s)}{s^2} = \frac{2}{3} \frac{f_V^2}{M^2}.
\ee
As a test of the expression for $\amuhvp$, 
using Eq.\ (109) in the review~\cite{Jegerlehner:2009ry}, we obtain 
\be
\amuhvp = \frac{8}{9}\alpha^2 f_V^2\frac{\mmu^2}{M^2} \hat K(M^2), 
\ee
where the kernel $\hat K(s)$ appropriate for the timelike region is
given in terms of elementary functions in \cite{Jegerlehner:2009ry}.  
Setting $M/\mmu=2m_{\pi^\pm}=2.64187$, we have
$\hat K(M^2)=0.63344$ and we thus obtain $\amuhvp=0.0806733\alpha^2
f_V^2$.  On the other hand, performing numerically the integral
(\ref{eq:GHred}) with the ${\cal H}_i$ given by
Eqs.\ (\ref{eq:f2final}--\ref{eq:f1final}) and the ${\cal G}_i$ read
off from Eq.\ (\ref{eq:GmunuModel}), we obtain the same value of
$\amuhvp$ to all indicated digits.

%  Eqs.\ (\ref{eq:master1}, \ref{eq:master2}) 
% and from the model correlator (\ref) 
% \be \amuhvp = (8\alpha^2/3)f_V^2 \frac{\mmu^2}{M^2} ....... \ee

\begin{figure}[t]
\includegraphics[width=.5\textwidth]{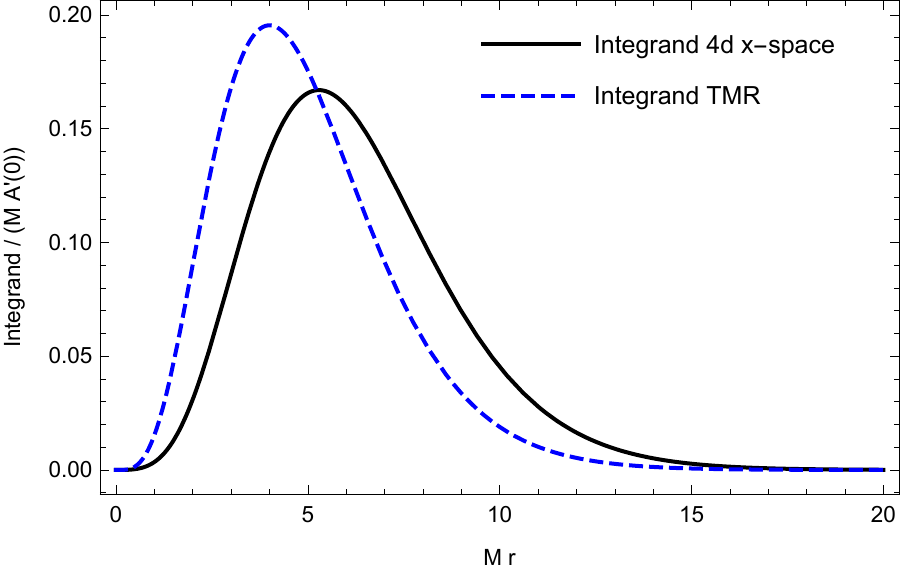}
\caption{The integrand in the integral over $|x|$ in Eq.\ (\ref{eq:Ap0}) to obtain the Adler function at the origin, ${\cal A}'(0)$, 
is displayed as a continuous curve.
For comparison, the dashed curved is the integrand in the integration over Euclidean time in the time-momentum representation;
see Eq.\ (\ref{eq:Dp0}) with ($r\doteq x_0$). In both cases, the model for the spectral function is given by Eq.\ (\ref{eq:rhomodel}) and
the displayed functions are normalized such that the area under the curves is unity. Note that for $M=800\,$MeV, $Mr=4$ corresponds to 1\,fm.}
\label{fig:intgnd}
\end{figure}

\subsection{Comparison with the time-momentum representation}

In order to prepare the discussion in the next section,
it is interesting to compare the derived formulae with the time-momentum representation,
in which only the spatial rotations are kept as manifest symmetries. 
The starting point in the TMR is the mixed-representation Euclidean correlator,
\be\la{eq:Gdef}
G(x_0)\delta_{k\ell} = - \int d^3\vec x\; G_{kl}(x),
\ee
which has the spectral representation~\cite{Bernecker:2011gh}
\be\la{eq:Gspecrep}
G(x_0) = \int_0^\infty d\omega\; \omega^2\rho(\omega^2)\; e^{-\omega |x_0|},
\qquad x_0\neq 0.
\ee
The vacuum polarization and the Adler function 
can be expressed as an integral over $G(x_0)$~\cite{Bernecker:2011gh,Francis:2013fzp},
\ba  \la{eq:PihatGt}
&&\Pi(Q_0^2)-\Pi(0)  
 = \\ && \qquad  
\int_{0}^\infty \!\!\! dx_0\, G(x_0)\Big[x_0^2 - \frac{4}{Q_0^2}\sin^2(\half Q_0 x_0) \Big],
\nonumber
\\ \la{eq:Adler}
&& {\cal A}(Q_0^2) \equiv  Q_0^2 \frac{d\,\Pi}{dQ_0^2} 
 = \frac{1}{Q_0^2} \int_0^\infty dx_0 \, G(x_0) 
\\ && \quad \qquad \qquad \left(2-2\cos(Q_0 x_0) - Q_0x_0\sin(Q_0x_0)\right).
\nonumber
\qquad 
\ea
In particular, the slope of the Adler function is given by 
\be\la{eq:Dp0}
% D'(0)=\lim_{{Q^2\to0}}\frac{D(Q^2)}{Q^2} = \pi^2 \int_0^\infty dx_0 \;x_0^4\, G(x_0).
{\cal A}'(0)=\lim_{{Q^2\to0}}\frac{{\cal A}(Q^2)}{Q^2} = \frac{1}{12} \int_0^\infty dx_0 \;x_0^4\, G(x_0).
\ee
The integrand for ${\cal A}'(0)$ is displayed as a dashed curve in Fig.\ \ref{fig:intgnd}.
Finally, the quantity $\amuhvp$ is given by~\cite{Bernecker:2011gh,DellaMorte:2017dyu}
\ba
a_\mu^{\rm HLO} &=& \Big(\frac{\alpha}{\pi}\Big)^2\int_0^\infty dx_0\;G(x_0) \; \tilde f(x_0),
\\ \la{eq:res}
\tilde f(x_0) &=&  \frac{2\pi^2}{m_\mu^2}
\Big(-2 + 8 \gamma_{\rm E} + \frac{4}{\hat x_0^2} - 2 \pi \hat x_0
\\ && + \hat x_0^2 - \frac{8}{\hat x_0} K_1( 2 \hat x_0) + 8 \log(\hat x_0)  +  8 I_p(\hat x_0)\Big),
\nonumber
\ea
where $\hat x_0 = m_\mu x_0$.
Here $\gamma_{\rm E}=0.577216..$ is Euler's constant and 
\be
I_p(z)=\frac{\pi z}{4}+ \frac{1}{8} G_{1,3}^{2,1}\left(z^2|
\begin{array}{c}
 \frac{3}{2} \\
 0,1,\frac{1}{2} \\
\end{array}
\right).
% \frac{\pi t}{4}+ \frac{1}{8} {\rm MeijerG}[\{\{3/2\}, \{\}\}, \{\{0, 1\}, \{1/2\}\}, t^2].
\ee
One aspect that is common between the CCS method and the TMR method is that, in continuum QCD,
the integrand for ${\cal A}'(Q^2)$, $\Pi(Q^2)-\Pi(0)$  and  $\amuhvp$ 
is of order $|x|$ (respectively order $x_0$) at short distances.
% This property implies that cutoff effects are of order $a^2$ if on-shell improvement is implemented.
% Indeed the counter-terms for full off-shell improvement would be local operators that vanish by the equations
% of motion and therefore are relevant only when the two vector current come close together.
% As we shall see shortly, if $G_{\mu\nu}(x)={\rm O}(|x|^{-6})$, the integrand at small $|x|$ is of order $|x|$.

\section{Lattice QCD aspects\la{sec:LQCD}}

In this section we discuss some of the possible implementation strategies
of the covariant coordinate-space method.
% \subsection{CCS method implementation strategies}
% We have seen that the integrand with the covariant position-space method looks similar to the integrand in the TMR method.
In the master relation (\ref{eq:master1}), the O(4) symmetry of the integrand in the continuum allows for a lot of 
flexibility when implementing an estimator for $\amuhvp$ in lattice QCD.
The finite lattice spacing as well as the toric boundary conditions break the O(4) symmetry. Therefore, 
for a number of classes $\Omega$ of lattice vectors on a lattice of dimensions $L^4$,
it is useful to investigate consistency and statistical precision of the estimators
\be\la{eq:amuOmega}
a_\mu^{\rm est}(\Omega) = 2\pi^2 \frac{|\varepsilon^{(.)}|}{|\Omega|} 
\sum_{k=1}^{|\Omega|}\sum_{n=1}^{n_{\rm max}} \Big[|x|^3\,G_{\mu\nu}(x) \,H_{\mu\nu}(x)\Big|_{x=n\varepsilon^{(k)}}
\ee
where the lattice vectors $\varepsilon^{(k)}$ belong to an orbit $\Omega$ of the hypercubic group
(for instance, $\Omega=\{a(\pm1,\pm1,\pm1,0),\;a(\pm1,\pm1,0,\pm1),\dots\}$, for which $|\Omega|=32$ and $|\epsilon^{(.)}|=a\sqrt{3}$.
Of course, a more sophisticated integration scheme may also be applied. If a $T\times L^3$ lattice is used ($T\neq L$),
the orbit is restricted to vectors related by the three-dimensional hypercubic symmetry group.

One may  wonder how the signal-to-noise will compare between a position-space and the momentum-space method
 in a lattice QCD simulation.
We argue that, in a large volume, one may expect an advantage with the covariant coordinate-space method.
The reason is simple and stems from the fact that the `signal', $G_{\mu\nu}(x)\sim e^{-M|x|}$, falls off faster than the
square-root of its variance $\sigma$. The latter is expected to drop only like $\sigma\sim e^{-m_\pi |x|}$
for the isovector contribution. In the case of disconnected diagrams, $\sigma$ is asymptotically independent of $|x|$
with the standard algorithm. 
With the proposed CCS method, one may sum in the variable $|x|$ up to a maximum $R$ 
(effectively performing a weighted average over all orbits),
\be\la{eq:amueff}
a_\mu^{\rm eff}(R) = a^4 \sum_{x:\,|x|<R} G_{\mu\nu}(x)\,H_{\mu\nu}(x),
\ee
with $|x|$ the Euclidean norm of the position vector $x$. The truncation distance $R$ is chosen so that the incurred
error is sufficiently small. In practice, an extrapolation to $R=\infty$ based on Eq.\ (\ref{eq:GmunuModel}),
or a more sophisticated version involving the two-pion continuum, may be used. 
The important point is that only those points in $x$ are included in the sum which contribute 
up to a certain precision. By contrast, in a momentum-space method, the input data $\Pi_{\mu\nu}(Q)$ 
already involves a sum over the correlator $G_{\mu\nu}(x)$ over the whole volume, even though 
points very distant from the origin end up barely contributing to $\amuhvp$.
We therefor expect the CCS method to be superior.

In the TMR method, the variable $x_0$ plays a role analogous to $|x|$
in the CCS method.  The sum in $x_0$ is truncated at some $x_0^{\rm
  max}$, beyond which the correlator is estimated using an
extrapolation based on its spectral representation. Although less
severe than in the four-momentum space method, the unfavorable aspect
that \emph{spatially} very distant points ($[\sum_{i=1}^3
  x_i^2]^{1/2}$ large) are included in the estimator for $\amuhvp$
remains present.  The CCS method should be at an advantage here,
particularly for disconnected diagram contributions.

Having made this point, there are other considerations that contribute
to choosing a computational method. One important consideration is the
control over finite-volume effects.  The nature of these corrections
are by now quite well understood in the TMR
method~\cite{DellaMorte:2017dyu}, although more direct numerical
studies (involving several volumes, all other parameters being held
fixed) are desirable.  It remains to be studied how large the
finite-size effects are in the CCS method. If the spatial torus size
is $L$, the sum (\ref{eq:amueff}) would have to be truncated at
$R_{\rm max}\leq L/2$.  However, one may be able to probe the
long-distance part of the correlator further, since one can choose for instance the orbit of
the vector $\varepsilon=a\cdot(1,1,1,1)$ and apply the estimator
(\ref{eq:amuOmega}) for distances beyond $R_{\rm max}$.  This
procedure effectively extends the distance reach by a factor of two,
at the cost of having less volume averaging.

% \section{Epilogue}

Finally, we remark that the framework presented could also be used to
calculate the transverse part of the axial-current correlator. In
particular, the explicit projection onto the transverse part via the
tensor $H_{\mu\nu}(x)$ takes care of removing the pion pole present in
the longitudinal channel.

\acknowledgments{This work was supported in part by DFG through CRC\,1044
\emph{The low-energy frontier of the Standard Model}.}

%%%%%%%%%%%%%%%%%%%%%%%%%%%%%%%%%%%%%%%%%%%%%%%%%%%%%%%%%%%%%%%%
\bibliography{/Users/harvey/BIBLIO/viscobib.bib}
%%%%%%%%%%%%%%%%%%%%%%%%%%%%%%%%%%%%%%%%%%%%%%%%%%%%%%%%%%%%%%%%

\end{document}